\documentclass[journal,twoside,web]{ieeecolor}
\usepackage{generic}
\usepackage{stfloats}
\usepackage{cite}
\usepackage{amsmath,amssymb,amsfonts}
\usepackage{algorithmic}
\usepackage{graphicx}
\usepackage{algorithm,algorithmic}
\usepackage{hyperref}
\usepackage{makecell} 
\hypersetup{hidelinks=true}
\usepackage{textcomp}
\usepackage{array} 

\def\BibTeX{{\rm B\kern-.05em{\sc i\kern-.025em b}\kern-.08em
    T\kern-.1667em\lower.7ex\hbox{E}\kern-.125emX}}
\begin{document}
\title{Anatomy-Informed Deep Learning and Radiomics for Automated Neurofibroma Segmentation in Whole-Body MRI}
\author{Georgii Kolokolnikov, Marie-Lena Schmalhofer, Lennart Well, Said Farschtschi, Victor-Felix Mautner, Inka~Ristow, and Ren\'e Werner
\thanks{This work was supported by DFG grant (DFG SPP 2177), project number 515277218 and the Bundesverband Neurofibromatose e.V. (Corresponding author: Georgii Kolokolnikov).}
\thanks{Georgii Kolokolnikov and Ren\'e Werner are with the Institute for Applied Medical Informatics, the Institute of Computational Neuroscience, and the Center for Biomedical Artificial Intelligence (bAIome), University Medical Center Hamburg-Eppendorf, 20246 Hamburg, Germany (e-mail: g.kolokolnikov@uke.de; r.werner@uke.de).}  
\thanks{Marie-Lena Schmalhofer, Lennart Well, and Inka Ristow are with the Department of Diagnostic and Interventional Radiology and Nuclear Medicine, University Medical Center Hamburg-Eppendorf, 20246 Hamburg, Germany (e-mail: m.schmalhofer@uke.de; l.well@uke.de; i.ristow@uke.de).} 
\thanks{Said Farschtschi and Victor-Felix Mautner are with the Department of Neurology, University Medical Center Hamburg-Eppendorf, 20246, Hamburg, Germany (e-mail: s.farschtschi@uke.de; v.mautner@uke.de).}}

\maketitle
\begin{abstract}
Neurofibromatosis Type 1 is a genetic disorder characterized by the development of neurofibromas (NFs), which exhibit significant variability in size, morphology, and  anatomical location. Accurate and automated segmentation of these tumors in whole-body magnetic resonance imaging (WB-MRI) is crucial to assess tumor burden and monitor disease progression. In this study, we present and analyze a fully automated pipeline for NF segmentation in fat-suppressed T2-weighted  WB-MRI,  consisting of three stages: anatomy segmentation, NF segmentation, and tumor candidate classification.  In the first stage, we use the MRSegmentator model to generate an anatomy segmentation mask, extended with a high-risk zone for NFs. This mask is concatenated with the input image as anatomical context information for NF segmentation.  The second stage employs an ensemble of 3D anisotropic anatomy-informed U-Nets to produce an NF segmentation confidence mask. In the final stage, tumor candidates are extracted from the confidence mask and classified based on radiomic features, distinguishing tumors from non-tumor regions and reducing false positives. We evaluate the proposed pipeline on three test sets representing different conditions: in-domain data (test set 1), varying imaging protocols and field strength (test set 2), and low tumor burden cases (test set 3). Experimental results show a 68\% improvement in per-scan Dice Similarity Coefficient (DSC), a 21\% increase in per-tumor DSC, and a two-fold improvement in F1 score for tumor detection in high tumor burden cases by integrating anatomy information. The method is integrated into the 3D Slicer platform for practical clinical use, with the code publicly accessible.
\end{abstract}


 \begin{figure}[t]
     \centering
     \includegraphics[width=0.75\linewidth]{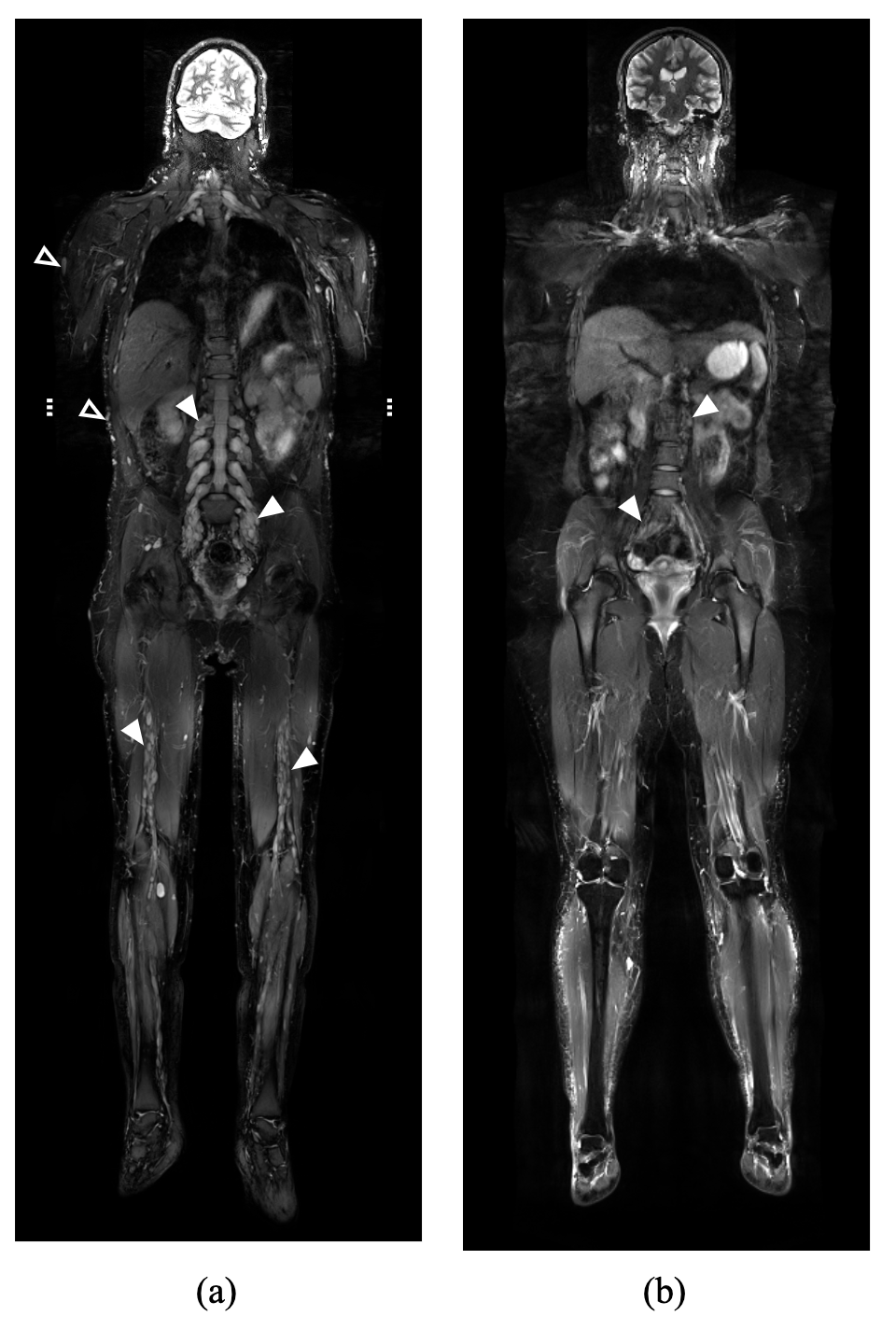}
     \caption{Fat-suppressed T2-weighted whole-body MRI of two patients with varying tumor burden. (a) 34-year-old male with a tumor burden of 1,591 cm$^{3}$, showing plexiform neurofibromas (PNFs, filled arrowheads) and subcutaneous neurofibromas (empty arrowheads). (b) 22-year-old female with a tumor burden of 96 cm$^{3}$,  showing PNFs (arrowhead).}
     \label{fig:nf}
 \end{figure}

\section{Introduction}
\label{sec:introduction}
\IEEEPARstart{N}{eurofibromatosis} Type 1 (NF1) is an autosomal-dominant genetic disorder affecting 1 in 2,500 to 3,000 individuals \cite{lammert_prevalence_2005}. The clinical manifestations of NF1 are diverse, with the hallmark feature being neurofibromas (NFs) distributed throughout the body \cite{friedman_type_1997}. NFs (Fig. \ref{fig:nf}), which can be cutaneous, subcutaneous, or plexiform (PNFs) \cite{thakur_multiparametric_2024}, vary in size, number, and location, leading to disease heterogeneity \cite{plotkin_quantitative_2012}. 
Compared to cutaneous and subcutaneous NFs, PNFs are of higher clinical importance due to the elevated risk of transforming into malignant peripheral nerve sheath tumors (MPNSTs). MPNSTs are a leading cause of premature mortality, with a poor five-year survival rate of 21\% \cite{evans_malignant_2002}. 

Whole-body magnetic resonance imaging (WB-MRI) is an essential tool for assessing tumor burden in NF1 patients \cite{ahlawat_current_2020}. Fat-suppressed T2-weighted (T2w) MR sequences are particularly useful for detecting NFs, as they appear as hyperintense lesions against the surrounding tissue \cite{chan_magnetic_2013}.

Traditional approaches to the analysis of WB-MRI of NF1 patients rely on manual tumor volume assessment, which is prone to inter- and intra-observer variability and becomes labor-intensive in high tumor burden cases \cite{plotkin_quantitative_2012, heffler_tumor_2017, weizman_interactive_2012, pratt_tumor_2015}. Advances in medical imaging and artificial intelligence have led to the development of promising automated and semi-automated segmentation methods. However, NF segmentation face specific significant challenges:

\begin{itemize}
    \item \textbf{Complex NF morphology}. Variability in NF appearance causes segmentation inconsistencies \cite{solomon_automated_2004, weizman_interactive_2012, pratt_tumor_2015}. This is, for example, reflected by a
    reported Dice Similarity Coefficient (DSC) of only 25\% for automated nnU-Net-based NF segmentation in 3D WB-MRI \cite{zhang_dins_2022}.
    \item \textbf{Limitations of interactive segmentation techniques}. Semi-automated methods, while reducing manual efforts, remain time-consuming, error-prone, and reliant on user input \cite{pratt_tumor_2015, ho_image_2020, zhang_dins_2022}. 
    \item \textbf{Scarcity of annotated data}. Only one out of eight \cite{weizman_interactive_2012, pratt_tumor_2015, solomon_automated_2004, zhang_dins_2022, ho_image_2020, weizman_pnist_2014, wu_deep_2020, wu_dh-gac_2022} reviewed NF segmentation studies included over 100 WB-MRI scans \cite{zhang_dins_2022}.
    \item \textbf{Lack of standardized evaluation}. Inconsistent evaluation protocols complicate method comparisons \cite{solomon_automated_2004, weizman_interactive_2012, weizman_pnist_2014, pratt_tumor_2015, wu_deep_2020, ho_image_2020, zhang_dins_2022, wu_dh-gac_2022}.
\end{itemize}

To address these challenges, we propose and comprehensively validate a novel fully-automated NF segmentation approach. Our key contributions include:

 \begin{enumerate}
     \item \textbf{Anatomy-informed NF segmentation}. To improve segmentation accuracy, we leverage segmentation masks of organs and add a zone with a high likelihood of NF occurrence (NF high-risk zone) to condition an ensemble of 3D anisotropic U-Nets.
     \item \textbf{Tumor candidate classification}. Focusing on high tumor burden cases, we present a random forest-based classification algorithm that integrates anatomical knowledge and radiomics to correct NF segmentation masks.
     \item \textbf{Pipeline integration into 3D Slicer}. We integrate the proposed segmentation pipeline into 3D Slicer \cite{fedorov_3d_2012}, a widely used medical imaging platform, facilitating its use in a clinical environment. 
 \end{enumerate}

 The segmentation pipeline (source code, trained models) is publicly available at \href{https://github.com/IPMI-ICNS-UKE/NFSegmentationPipeline}{https://github.com/IPMI-ICNS-UKE/NFSegmentationPipeline}. This paper includes a supplementary video demonstrating the proposed NF segmentation pipeline in 3D Slicer. It showcases segmentation and post-processing modes on two fat-suppressed T2w  WB-MRI cases. The video is in MP4 format (H.264 codec, 1080p resolution).

\section{Related Work}
\label{sec:related work}

Early work on NF segmentation relied on semi-automated techniques. In 2004, Solomon \textit{et al.} \cite{solomon_automated_2004} introduced a thresholding-based method with manual initialization, which was prone to errors in iso-intense regions. 

In 2012, Weizman \textit{et al.} \cite{weizman_interactive_2012} proposed a method with manual initialization, followed by a segmentation based on tumor connectivity. It struggled with intensity similarities between PNFs and healthy tissues. A refinement in 2013 utilized a histogram-based tool requiring minimal user input \cite{weizman_pnist_2014}; however, it showed lower accuracy for small-volume tumors and diffusion of segmentation into high-signal-intensity structures adjacent to tumors.

A shift towards deep learning (DL) in NF segmentation was marked by Wu \textit{et al.} in 2020 \cite{wu_deep_2020}, who introduced the Deep Parametric Active Contour Model (Deep-PAC). It combined active contour models with DL features. Deep-PAC demonstrated superior performance over Fully Convolutional Networks (FCN) and U-Net. However, it required manual initialization and was computationally intensive. 

Advancements in fully automated methods were demonstrated by Ho \textit{et al.} in 2020 \cite{ho_image_2020}. They developed a multi-spectral neural network classifier that utilized diffusion-weighted imaging data. Although this method reduced intra- and inter-observer variability, it struggled with misclassifying normal anatomy as pathological. 

In 2021, Zhang \textit{et al.} \cite{zhang_dins_2022} proposed Deep Interactive Networks (DINs), using an Exponential Distance Transform to convert user interactions into guide maps. Compared to 3D U-Net and nnU-Net, DINs achieved significantly higher DSC, though it required user interaction.

In 2022, Wu \textit{et al.} introduced a Deep Hybrid Contextual Feature Network, integrated with a Multi-Gradient Active Contour model \cite{wu_dh-gac_2022}. The method consistently outperformed both U-Net and FCN, but it faced challenges due to intensity inhomogeneity and fuzzy boundaries, in addition to numerical instabilities during contour evolution. 

Despite progress in NF segmentation, existing methods remain sensitive to tumor appearance variability, are often restricted to 2D operations, or are limited to segmenting only a few tumors within a region of interest. Our approach addresses these limitations by integrating anatomical knowledge and radiomics in a fully automated multistage pipeline. 

\section{Methods}
\label{sec:methods}

This single-center retrospective study was approved by the local ethics committee (2022-300201-WF, 2022-300201\_1-WF) with a waiver of informed consent, adhering to data protection regulations and the Declaration of Helsinki.

\begin{figure*}
    \centering
    \includegraphics[width=1\linewidth]{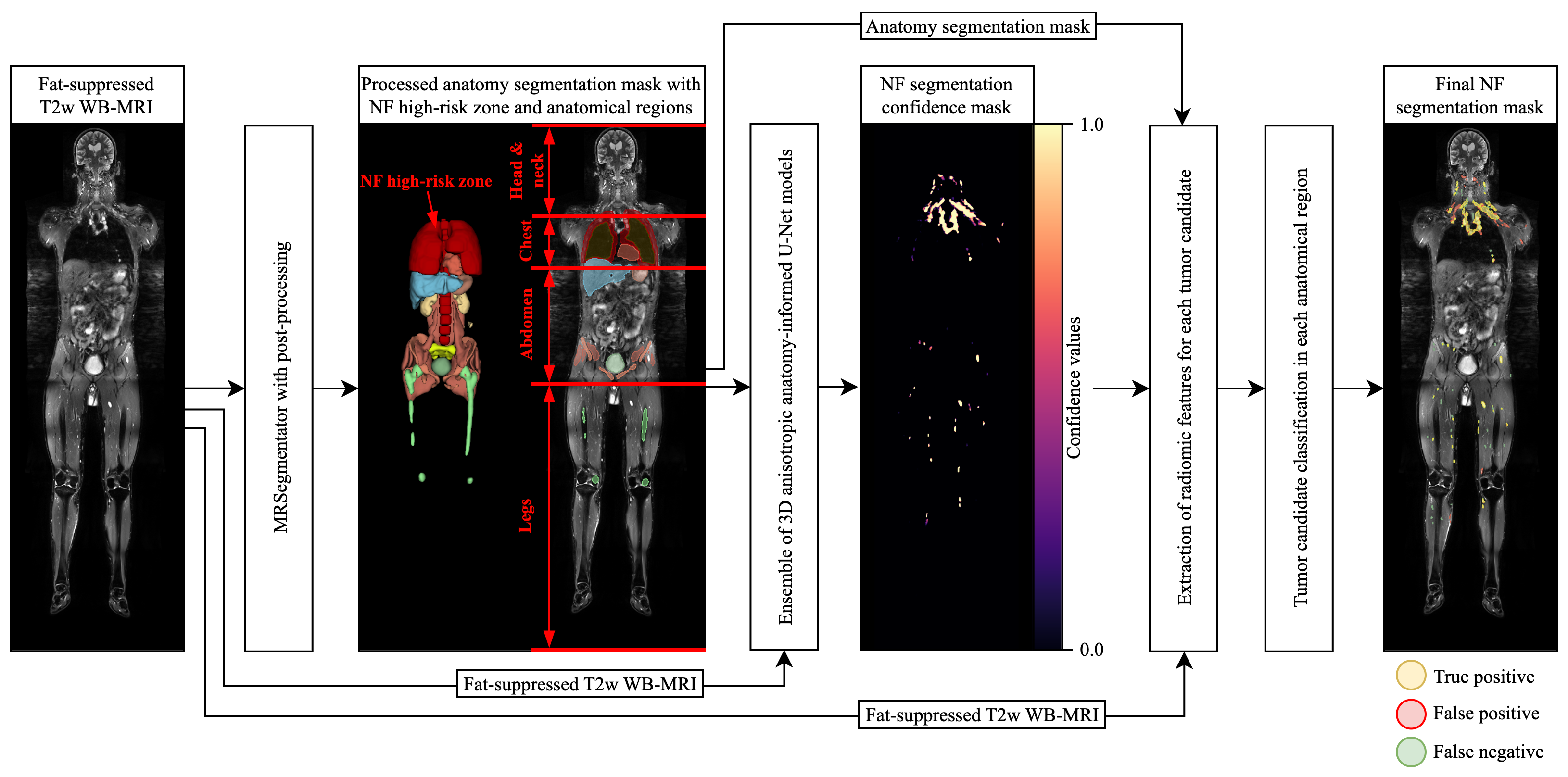}
    \caption{Proposed pipeline for automated neurofibroma (NF) segmentation, consisting of three main stages: (1) The input T2-weighted whole-body MRI (T2w WB-MRI) is segmented using the MRSegmentator model, followed by post-processing to refine the anatomy segmentation mask. (2) The input T2w WB-MRI and the refined anatomy segmentation mask are processed by an ensemble of 3D anisotropic anatomy-informed U-Net models to produce an NF segmentation confidence mask. (3) The input T2w WB-MRI, the anatomy segmentation mask, and the NF confidence mask are used to extract tumor candidates with their radiomic features. The tumor candidates are grouped by anatomical regions (head and neck, chest, abdomen, legs) and classified using random forest classifiers. The final NF segmentation mask is formed by filtering the tumor candidates. }
    \label{fig:pipeline}
\end{figure*}

The proposed novel fully-automated pipeline for NF segmentation in T2w WB-MRI contains three main stages (Fig. \ref{fig:pipeline}): anatomy segmentation using the MRSegmentator \cite{hantze_mrsegmentator_2024} (Section \ref{subsec:stage 1}), NF segmentation using an ensemble of 3D anisotropic anatomy-informed U-Nets (Section \ref{subsec:stage 2}), and tumor candidate classification leveraging radiomic features (Section \ref{subsec:stage 3}). Since the pipeline does not require user interaction, it reduces the manual workload and potential bias typically present in semi-automated methods \cite{solomon_automated_2004, weizman_interactive_2012, weizman_pnist_2014, pratt_tumor_2015, wu_deep_2020, zhang_dins_2022}. Furthermore, the pipeline can serve as a pre-segmentation step for refinement by interactive techniques \cite{nguyen_deepedit_2022, zhang_dins_2022}, though this is beyond the scope of the current study.

\subsection{Stage 1: Anatomy Segmentation Using MRSegmentator}
\label{subsec:stage 1}
\subsubsection{Rationale}

NFs are more likely to occur in certain anatomical regions, particularly along the nerves in the head, neck, trunk, and extremities, with a higher concentration around areas such as the spine, brachial plexus, and sciatic nerves \cite{staser_pathogenesis_2012, palmer_analysis_2004, plotkin_quantitative_2012, thakur_multiparametric_2024}. To enhance NF segmentation accuracy, we used anatomical context to focus the neural network model on areas with a higher likelihood of NFs (see Section \ref{subsec:stage 2}). This idea is inspired by the prompting strategy in the Segment Anything Model \cite{kirillov_segment_2023}. There, the prompts helped the neural network identify the region of interest.

\begin{figure}
    \centering
    \includegraphics[width=1\linewidth]{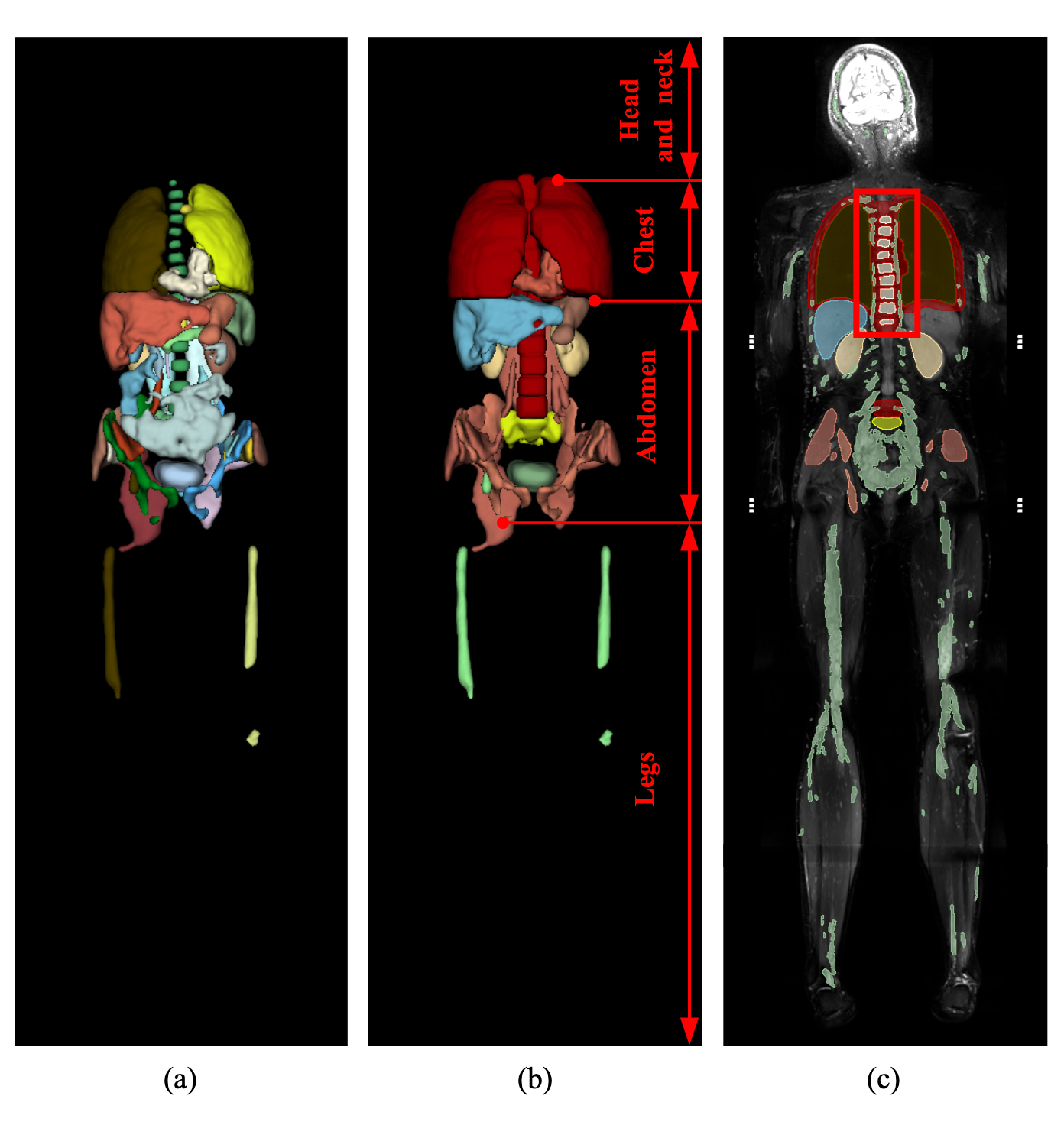}
    \caption{Comparison of initial and refined anatomy segmentation masks: (a) Initial coarse anatomy segmentation mask generated by MRSegmentator. (b) Refined mask with redundant anatomies removed, labels merged, neurofibroma (NF) high-risk zone added (red structure around the lungs and spine), and anatomical regions (red arrows) defined by anatomical landmarks: the highest and the lowest points of the lungs, and the lowest point of the hips (red dots). (c) T2-weighted whole-body MRI with updated anatomy segmentation masks and NF ground truth (green), highlighting NFs within the NF high-risk zone (red rectangle).}
    \label{fig:anatomical knowledge}
\end{figure}

\subsubsection{Prior Anatomical Knowledge}
To extract an anatomy segmentation mask from the input data, we utilized MRSegmentator \cite{hantze_mrsegmentator_2024}. Despite its capabilities, the model exhibited limitations when applied to the highly anisotropic fat-suppressed T2w WB-MRIs of this study due to low resolution in certain planes and low contrast of soft tissues. These issues contributed to less reliable segmentation outputs from MRSegmentator as shown in Fig. \ref{fig:anatomical knowledge}.a, e.g., in structures like the duodenum, small bowel, colon, and bones.

Therefore, the anatomy segmentation was refined by post-processing and incorporating anatomical knowledge of typical NF locations \cite{staser_pathogenesis_2012, palmer_analysis_2004, plotkin_quantitative_2012, thakur_multiparametric_2024}. Organs with DSC below 0.85 on the NAKO dataset \cite{hantze_mrsegmentator_2024} and vessels were excluded due to poor segmentation performance. Additionally, paired organs were merged, and muscle groups were consolidated into one label (Table \ref{tab:anatomy_segmentation_mask}).

\begin{table}
    \centering
\caption{Anatomy Segmentation Mask Construction Based on Reported MRSegmentator Performance and Knowledge of Typical Neurofibroma (NF) Localizations}
\label{tab:anatomy_segmentation_mask}
    \begin{tabular}{|>{\centering\arraybackslash}p{0.2\linewidth}|>{\centering\arraybackslash}p{0.08\linewidth}|>{\centering\arraybackslash}p{0.1\linewidth}|>{\centering\arraybackslash}p{0.15\linewidth}|>{\centering\arraybackslash}p{0.18\linewidth}|} \hline 
           Organs&  DSC&  NF can be inside&  NF can be in proximity& Included in the mask\\ \hline 
           Heart&  0.95&  No&  Yes& Yes\\ \hline 
           Lungs&  0.97&  No&  Yes& Yes (merged)\\ \hline 
           Esophagus&  0.84&  No&  Yes& No\\ \hline 
           Liver&  0.95&  No&  No& Yes\\ \hline 
           Spleen&  0.89&  No&  No& Yes (merged into stomach)\\ \hline 
           Pancreas&  0.75&  No&  No& No\\ \hline 
           Gallbladder&  0.74&  No&  No& No\\ \hline 
           Stomach&  0.89&  No&  No& Yes\\ \hline 
           Duodenum&  0.76&  No&  No& No\\ \hline 
 Small bowel& 0.75& No& Yes&No\\ \hline 
 Colon& 0.81& No& Yes&No\\ \hline 
 Kidneys& 0.88 / 0.90& No& Yes&Yes (merged)\\ \hline 
 Adrenal glands& 0.65 / 0.61& No& Yes&No\\ \hline 
 Urinary bladder& 0.95& No& Yes&Yes\\ \hline 
 Spine& 0.89& Yes& Yes&Yes\\ \hline 
 Sacrum& 0.87& Yes& Yes&Yes\\ \hline 
 Hips& 0.89 / 0.89& No& Yes&Yes (merged)\\ \hline 
 Femurs& 0.94 / 0.93& No& Yes&Yes (merged)\\ \hline 
 Gluteus maximus muscles& 0.85 / 0.91& Yes& Yes&Yes (merged into muscles)\\ \hline 
 Gluteus medius muscles& 0.92 / 0.90& Yes& Yes&Yes (merged into muscles)\\ \hline 
 Gluteus minimus muscles& 0.88 / 0.89& Yes& Yes&Yes (merged into muscles)\\ \hline 
 Autochthonous muscles& 0.94 / 0.94& Yes& Yes&Yes (merged into muscles)\\ \hline 
 Iliopsoas muscles& 0.92 / 0.91& Yes& Yes&Yes (merged into muscles)\\ \hline 
 Aorta& 0.90& No& Yes&No\\ \hline 
 Inferior vena cava& 0.81& No& Yes&No\\ \hline 
 Portal/splenic vein& 0.54& No& Yes&No\\ \hline 
 Iliac arteries& 0.77 / 0.72& No& Yes&No\\ \hline 
 Iliac veins& 0.80 / 0.82& No& Yes&No\\ \hline
 \multicolumn{5}{p{230pt}}{DSC – reported Dice Similarity Coefficient of MRSegmentator on the NAKO Dataset \cite{hantze_mrsegmentator_2024}.}\\  
    \end{tabular}
\end{table}

Based on prior knowledge about typical NF localizations \cite{staser_pathogenesis_2012, palmer_analysis_2004, plotkin_quantitative_2012, thakur_multiparametric_2024}, we defined an NF high-risk zone near the lungs (around ribs and intercostal nerves) and the spine (near spinal nerves). The NF high-risk zone, i.e. areas with a high probability of the occurrence of NFs, was added as a separate label to the anatomy segmentation mask by dilating segmentation masks of the lungs and spine.

The refined mask (Fig. \ref{fig:anatomical knowledge}.b) included 11 organ labels and an additional label for the NF high-risk zone. Fig. \ref{fig:anatomical knowledge}.c shows NFs within the labeled NF high-risk zone. We also defined specific anatomical regions based on landmarks:

\begin{itemize}
    \item \textbf{Head and neck}. Above the highest point of the lungs.
    \item \textbf{Chest}. Between the lungs' lowest and highest points.
    \item \textbf{Abdomen}. Between the lowest point of the lungs and the lowest point of the hips (ischium).
    \item \textbf{Legs}. Below the lowest point of the hips.
\end{itemize}

\subsection{Stage 2: Neurofibroma Segmentation Using an Ensemble of 3D Anisotropic Anatomy-Informed U-Nets}
\label{subsec:stage 2}

\subsubsection{Anisotropy of NF Data}
The T2w WB-MRI data are highly anisotropic, with the anterior-posterior axis voxel size being over ten times larger than for the cranio-caudal and lateral axes. 
We hypothesized that maintaining this anisotropic spacing during processing can improve segmentation accuracy by avoiding interpolation artifacts. We, therefore, resampled the data to a unified mean anisotropic spacing derived from the dataset for consistency (see Section \ref{subsubsec:demographic characteristics}).

\subsubsection{Segmentation Model Architecture}
We selected U-Net as the backbone architecture for its versatility. Unlike data-hungry transformer-based models (e.g., UNETR, Swin \mbox{UNETR}) \cite{hatamizadeh_unetr_2021, crimi_swin_2022}, U-Net is more suitable for datasets with limited samples, like in the NF segmentation case. 

Our approach differs from previous methods \cite{solomon_automated_2004, weizman_interactive_2012, pratt_tumor_2015, ho_image_2020, wu_deep_2020, zhang_dins_2022, wu_dh-gac_2022} by leveraging anatomical context to enhance NF segmentation accuracy. We concatenated the anatomy segmentation mask with the T2w WB-MRI as a two-channel input to guide the neural network to focus on specific regions within the T2w WB-MRI. While multiple methods exist to integrate anatomical knowledge into neural networks \cite{nguyen_deepedit_2022, zhang_dins_2022, kirillov_segment_2023, ahlawat_current_2016}, we opted for this simpler concatenation method to minimize architectural changes to the baseline U-Net model. This allowed us to more directly assess the impact of anatomical knowledge on segmentation performance.

To handle anisotropy, the 3D anisotropic U-Net maintained the original voxel resolution. With a patch size of 10 x 640 x 256, it balanced computational efficiency and anatomical context. The nnU-Net framework \cite{isensee_nnu-net_2021} was applied to optimize the architecture, featuring seven encoder and six decoder stages with convolutional blocks, instance normalization, and leaky rectified linear unit activation. Strides and kernel sizes were adapted to downsample along cranio-caudal and lateral axes while preserving anterior-posterior resolution. A 1x1x1 convolution with sigmoid activation mapped tumor probability.

\subsubsection{Ensembling}
We employed an ensemble of three 3D anisotropic anatomy-informed U-Nets, each trained with random initialization and different data splits via three-fold cross-validation. The tumor probability masks of the individual models were averaged to reduce variance and bias in predictions, generating an NF segmentation confidence mask.

\subsection{Stage 3: Tumor Candidate Classification}
\label{subsec:stage 3}
\subsubsection{Rationale}
The tumor candidate classification step was introduced to improve specificity in NF segmentation, i.e., the reduction the number of typical misclassification of non-tumorous tissues like muscles and fat as NFs. 

\subsubsection{Extracting Tumor Candidates}
After generating the NF segmentation confidence mask, thresholding was applied to distinguish tumors from the background. The confidence threshold, a hyperparameter, balanced true tumor detection against false positives. A higher confidence threshold (0.5 - default value) minimized false positives but could miss some true tumors (false negatives), whereas a lower confidence threshold (0.25 - median of the lowest confidence values in the predicted NF segmentation masks) captured more true tumors but increased false positives. Based on the resulting binary mask, component analysis was used to identify tumor candidates, grouped by anatomical region to account for appearance variability.

\subsubsection{Extracting Radiomic Features}
For each tumor candidate, we extracted 800 radiomic features (shape, intensity, texture) using PyRadiomics \cite{van_griethuysen_computational_2017}, leveraging their potential for NF tumor classification \cite{ristow_evaluation_2022, jansma_preoperative_2024}. Feature selection followed standard radiomics workflow \cite{lavrova_review_2024}: removal of features with near-zero variance, retention of non-intercorrelated features using Spearman correlation, and selection of the top 10 features via recursive feature elimination with a random forest classifier.

\subsubsection{Classification}
Selected features describing each tumor candidate were classified using region-specific random forest classifiers to account for NF variability across anatomical regions. Random forests were chosen as classifiers for their robustness and ability to handle small datasets and high-dimensional feature spaces \cite{breiman_random_2001}. Identified non-tumor candidates were excluded, forming the final NF segmentation mask.

\section{Experiments}
\label{sec:experiments}
\subsection{Data Description}
\label{subsec:data description}

\subsubsection{Source of Data}
\label{subsubsec:source of data}
The NF dataset was acquired at the neurofibromatosis outpatient clinic at the University Medical Center Hamburg-Eppendorf, using a Siemens Magnetom scanner (Siemens Healthineers, Erlangen, Germany). The inclusion criteria were: confirmed NF1 diagnosis \cite{noauthor_neurofibromatosis_1988}, no ongoing medication, and availability of WB-MRIs with documented peripheral nerve sheath tumors. T2w WB-MRIs were collected over 14 years (2006–2020), and tumor segmentation was manually performed by two radiologists (I.R. and M.-L. S.).

\subsubsection{Demographic, Imaging, and Clinical Characteristics}
\label{subsubsec:demographic characteristics}
The NF dataset included 109 T2w WB-MRI scans from 74 patients (37 males, 37 females; mean age: 29.7 years). The scans had high anisotropy (in-plane resolution of 0.63 mm x 0.63 mm; mean slice thickness of 7.79 mm). Ground truth annotations focused on PNFs due to their elevated risk of malignant transformation. A summary of the demographic and imaging characteristics of the data is presented in Table \ref{tab:demographic}.

\begin{figure}
    \centering
    \includegraphics[width=1\linewidth]{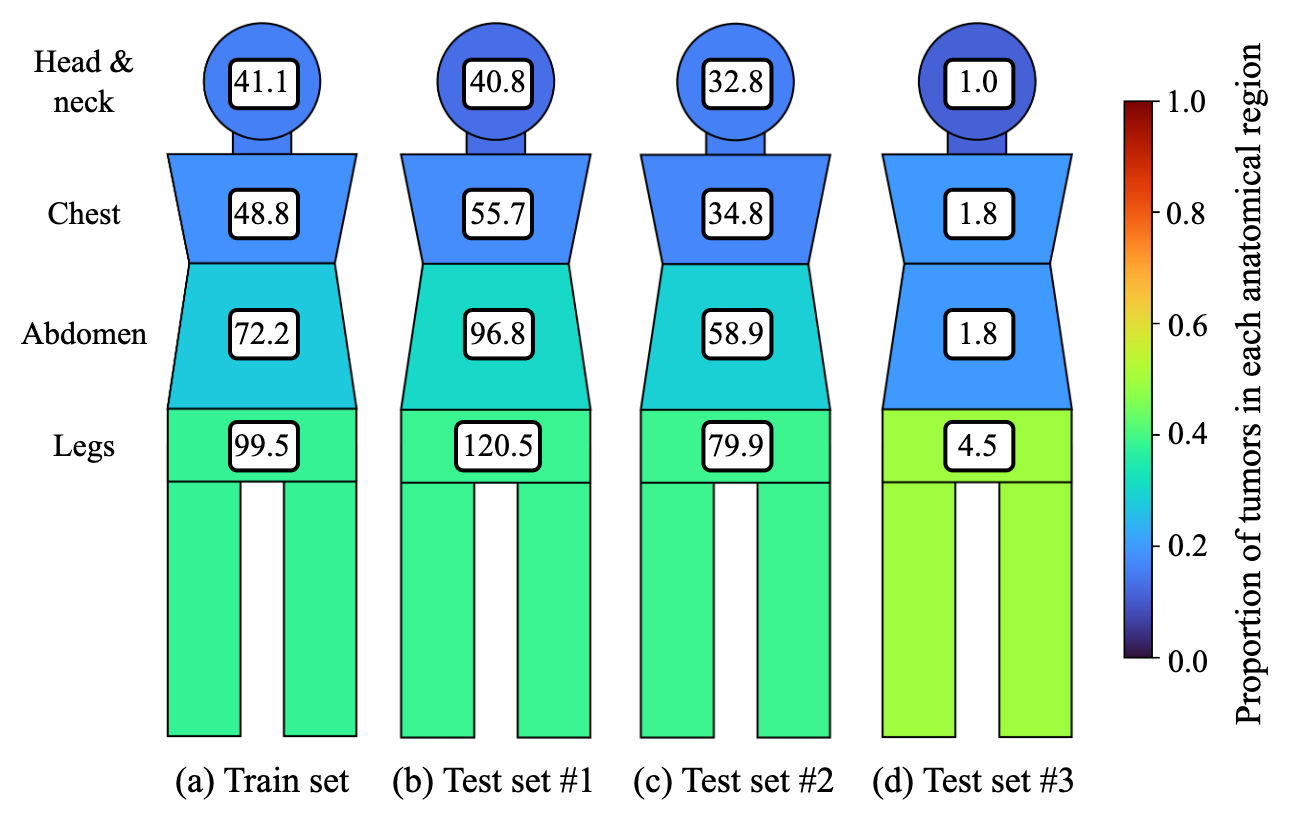}
    \caption{Distribution of tumors across anatomical regions in four datasets: (a) Train set; (b) Test set \#1; (c) Test set \#2; (d) Test set \#3. Each dataset is represented by a stylized human figure divided into anatomical regions. The numerical values within each region indicate the average absolute number of tumors in that region. The color of each region represents the proportion of tumors in that region relative to the total tumor count in each dataset, as indicated by the color bar.}
    \label{fig:tumor_count}
\end{figure}

Tumor burden and count varied across the patients with differences in anatomical distribution (Fig. \ref{fig:tumor_count}). The abdomen and legs were the most affected regions, displaying the highest tumor volumes and counts, whereas the chest, head, and neck were less involved, consistent with previous findings \cite{plotkin_quantitative_2012}.

\begin{table}
    \centering
\caption{Demographic and Imaging Characteristics of the Neurofibroma Dataset }
\label{tab:demographic}
    \begin{tabular}{|>{\centering\arraybackslash}p{0.4\linewidth}|>{\centering\arraybackslash}p{0.42\linewidth}|} \hline 
         Parameter& Value\\ \hline 
         \makecell[l]{T2w WB-MRI scans, total (n)}& 109\\ \hline 
         Train set& 63\\ \hline 
        Test set \#1& 13\\ \hline 
        Test set \#2& 11\\ \hline 
        Test set \#3& 22\\ \hline 
        \makecell[l]{Patients, total (n)}& 74\\ \hline 
        Train set& 42\\ \hline 
        Test set \#1& 13\\ \hline 
        Test set \#2& 9\\ \hline
Test set \#3&10\\\hline
\makecell[l]{Sex, total (m / f)}&37 / 37\\\hline
Train set&21 / 21\\\hline
Test set \#1&8 / 5\\\hline
Test set \#2&3 / 6\\\hline
Test set \#3&5 / 5\\\hline
 \makecell[l]{Mean age at scanning, \\ total (years, mean $\pm$ SD)}&29.7 $\pm$ 14.3\\\hline
Train set&31.9 $\pm$ 13.4\\\hline
Test set \#1&32.5 $\pm$ 13.6\\\hline
Test set \#2&27.7 $\pm$ 13.6\\\hline
Test set \#3&22.4 $\pm$ 15.7\\\hline
\makecell[l]{Imaging characteristics}&\makecell{3T and 1.5T, \\8 different MRI protocols}\\\hline
Train set&3T, 3 different MRI protocols\\\hline
Test set \#1&3T, 2 different MRI protocols\\\hline
Test set \#2&1.5T, 5 different MRI protocols\\\hline
Test set \#3&3T, 2 different MRI protocols\\\hline
\multicolumn{2}{p{210pt}}{T2w WB-MRI – T2-weighted whole-body MRI, SD – standard deviation, m – male, f – female.}\\  
    \end{tabular}
\end{table}

\subsubsection{Dataset Partitioning}
\label{subsubsec:dataset partitioning}
To evaluate the NF segmentation performance in different scenarios, the NF dataset was split into a train and three distinct test sets.  The partitioning strategy was as follows: patients scanned with a 1.5T MRI were allocated to Test Set \#2 (out-of-domain). For the remaining 3T scans, the tumor burden distribution was analyzed, and cases in the first quartile (tumor burden $<$47 cm$^{3}$) were assigned to Test Set \#3 (low tumor burden). The remaining cases were randomly split between the train and Test Set \#1 (in-domain) at the patient level. A summary of each set is provided below:

\begin{itemize} 
\item \textbf{Train set}. Showed a wide range of tumor burdens (median: 618.2 cm$^{3}$; interquartile range (IQR): 1276.9 cm$^{3}$) and counts (median: 224 tumors per scan; IQR: 319). The abdomen had the highest tumor burden (median volume: 219.9 cm$^{3}$; IQR: 690.8 cm$^{3}$; median count: 72.2), and the legs had the highest tumor count (median volume: 115.2 cm$^{3}$; IQR: 306.7 cm$^{3}$; median count: 99.5).

\item \textbf{Test set \#1 (in-domain)}. Included scans that matched the train set in terms of MRI protocols, field strength, and tumor burden to assess the model performance under known conditions. This set had the highest tumor burden (median: 1311.2 cm$^{3}$; IQR: 3872.1 cm$^{3}$) and showed significant variability in tumor counts (median: 62; IQR: 608). Tumor burden and variability were high across all regions, especially in the legs (median volume: 490.8 cm$^{3}$; IQR: 921.9 cm$^{3}$; median count: 120.5).

\item \textbf{Test set \#2 (out-of-domain)}. Included scans acquired with different MRI protocols (repetition time, echo time, flip angle) and at a lower field strength (1.5T compared to 3T used in the train set) to assess the model generalization ability across varying imaging environments. This set demonstrated moderate tumor burden (median: 442.4 cm$^{3}$; IQR: 512.1 cm$^{3}$) and count (median: 112; IQR: 288), with the abdomen (median volume: 139.3 cm$^{3}$; IQR: 145.9 cm$^{3}$; median count: 58.9) and legs (median volume: 97.4 cm$^{3}$; IQR: 227.9 cm$^{3}$; median count: 79.9) being the most affected regions.

\item \textbf{Test set \#3 (low tumor burden)}. Included scans from patients with low tumor burden (median: 5.6 cm$^{3}$; IQR: 10.9 cm$^{3}$) and count (median: 2; IQR: 5), allowing assessment of model sensitivity to sparse tumors. Tumor burden was minimal across anatomical regions.

\end{itemize}

\subsection{Experimental Design}
\label{subsec:experimental design}
The segmentation performance of the proposed pipeline was evaluated for the described three test sets. 
First, we compared a baseline 3D isotropic U-Net (input patch size: 128 x 128 x 128; spacing: 1.0 mm x 1.0 mm x 1.0 mm) with the 3D anisotropic U-Net (input patch size: 10 x 640 x 256; spacing: 7.8 mm x 0.625 mm x 0.625 mm) to assess the impact of anisotropic image processing. Next, we evaluated the benefit of anatomical knowledge integration by comparing the 3D anisotropic U-Net with a 3D anisotropic anatomy-informed U-Net that used two-channel input (T2w WB-MRI and anatomy segmentation mask). Finally, we tested the effect of tumor candidate classification by comparing an ensemble of 3D anisotropic anatomy-informed U-Nets with and without this post-processing step, employing both low- and high-confidence thresholds.

\subsection{Evaluation Metrics}
\label{subsec:evaluation metrics}
We used several metrics to evaluate NF segmentation performance, including DSC, Volume Overlap Error (VOE), and Absolute Relative Volume Difference (ARVD). DSC, a widely used metric for assessing spatial overlap \cite{isensee_nnu-net_2021}, was calculated per scan and per tumor. VOE and ARVD, which quantify volumetric differences, were included for consistent comparison with the previous study \cite{zhang_dins_2022}.

While DSC, VOE, and ARVD measured the accuracy of tumor segmentation, the F1 score was used to assess the NF detection performance per scan and anatomical region. This metric required calculating True Positives (TP), False Positives (FP), and False Negatives (FN). A tumor instance from a ground truth mask was counted as a TP if it overlapped with a prediction, and as an FN if it did not. A tumor instance from a predicted segmentation mask was counted as an FP if it did not match any ground truth.

\subsection{Implementation Details}
\label{subsec:implement details}
The NF segmentation pipeline was implemented in Python 3.9 and integrated into 3D Slicer \cite{fedorov_3d_2012} via MONAI Label. U-Net models were built using MONAI \cite{cardoso_monai_2022} and PyTorch, with architecture refinement via the nnU-Net framework. Image processing used SimpleITK, and anatomy segmentation relied on the MRSegmentator repository. Tumor candidate classification utilized PyRadiomics for feature extraction, Scikit-Learn for feature selection and random forest-based classification. All experiments were run on two NVIDIA RTX A6000 GPUs.

The 3D anisotropic anatomy-informed U-Net model had a batch size of 2 with 2-channel inputs (T2w WB-MRI and anatomy segmentation mask). We applied Z-score normalization to T2w WB-MRI images to standardize the intensity distribution across different scans, compensating for variability in MRI protocols and acquisition parameters. The anatomy segmentation masks were rescaled to the range 0-1, maintaining numerical stability during training. The model was trained with generalized Dice Focal Loss for imbalanced segmentation, and optimized with the AdamW algorithm (weight decay = $5 \times 10^{-5}$). An initial learning rate of $1 \times 10^{-4}$ was decayed by 0.99 per epoch over 1000 epochs, with the best validation checkpoint used as the final model. We used 3-fold cross-validation. Inference was performed using a sliding window approach with a 25\% overlap. 

\begin{table*}[t]
\caption{Segmentation Performance of Six Methods Across Three Test Sets. Methods: (1) 3D Isotropic U-Net, (2) 3D Anisotropic U-Net, (3) 3D Anisotropic Anatomy-Informed U-Net, (4) Ensemble of 3D Anisotropic Anatomy-Informed U-Nets, (5, 6) Ensemble of 3D Anisotropic Anatomy-Informed U-Nets with Low and High-Confidence Tumor Candidate Classification, Respectively. }
    \label{tab:general segmentation performance} 
    \centering
    \begin{tabular}{|c|c|c|c|c|c|c|} \hline 
         Method $i$&  1&  2&  3&  4&  5& 6\\ \hline 
         \makecell[l]{\textbf{Test set \#1}} & \multicolumn{6}{c|}{}\\ \hline 
         Per-scan DSC (Mean $\pm$ SD) $\uparrow$&  0.38 $\pm$ 0.23&  0.54 $\pm$ 0.23&  0.60 $\pm$ 0.17&  0.62 $\pm$ 0.16&  \textbf{0.64 $\pm$ 0.13}& 0.63 $\pm$ 0.15\\ \hline 
         $p$-value (Method $i$ vs $i-1$)&  -&  $<0.01$&  $<0.01$&  0.09&  0.87& 0.62
\\ \hline 
 Per-tumor DSC (Mean $\pm$ SD) $\uparrow$& 0.71 $\pm$ 0.09& 0.78 $\pm$ 0.09& 0.80 $\pm$ 0.14& 0.79 $\pm$ 0.12& 0.81 $\pm$ 0.12&\textbf{0.86 $\pm$ 0.10}\\ \hline 
 $p$-value (Method $i$ vs $i-1$)& -& $<0.01$& 0.23& 0.34& 0.06&$<0.01$\\ \hline 
         \makecell[l]{\textbf{Test set \#2}} & \multicolumn{6}{c|}{}\\ \hline 
         Per-scan DSC (Mean $\pm$ SD) $\uparrow$&  0.23 $\pm$  0.15&  0.48 $\pm$  0.16&  0.50 $\pm$  0.12&  \textbf{0.51 $\pm$  0.14}&  0.50 $\pm$  0.13& \textbf{0.51 $\pm$ 0.15}\\ \hline 
 $p$-value (Method $i$ vs $i-1$)& -& $<0.01$& 0.97& $<0.01$& $<0.01$&$<0.01$\\ \hline 
 Per-tumor DSC (Mean $\pm$ SD) $\uparrow$& 0.75 $\pm$ 0.22& 0.73 $\pm$ 0.20& 0.79 $\pm$ 0.12& 0.77 $\pm$ 0.20& 0.78 $\pm$ 0.22&\textbf{0.86 $\pm$ 0.11}\\ \hline 
 $p$-value (Method $i$ vs $i-1$)& -& 0.52& 0.02& 0.76& 0.68&$<0.01$\\ \hline 
 \makecell[l]{\textbf{Test set \#3}} & \multicolumn{6}{c|}{}\\ \hline 
 Per-scan DSC (Mean $\pm$ SD) $\uparrow$& 0.02 $\pm$ 0.01& 0.16 $\pm$ 0.18& 0.22 $\pm$ 0.23& \textbf{0.23 $\pm$ 0.24}& 0.20 $\pm$ 0.22&\textbf{0.23 $\pm$ 0.24}\\ \hline 
 $p$-value (Method $i$ vs $i-1$)& -& $<0.01$& $<0.01$& $<0.01$& $<0.01$&$<0.01$\\ \hline 
 Per-tumor DSC (Mean $\pm$ SD) $\uparrow$& 0.85 $\pm$ 0.19& 0.83 $\pm$ 0.19& 0.85 $\pm$ 0.18& 0.81 $\pm$ 0.24& 0.88 $\pm$ 0.14&\textbf{0.89 $\pm$ 0.18}\\ \hline 
 $p$-value (Method $i$ vs $i-1$)& -& 0.52& 0.04& 0.05& $<0.01$&0.54\\ \hline
 \multicolumn{7}{p{490pt}}{DSC – Dice Similarity Coefficient, SD – standard deviation. The highest metric values are in bold. Statistical significance was evaluated using the Wilcoxon signed-rank test for paired per-scan DSC between consecutive methods (Method $i$ vs. Method $i-1$), with Bonferroni correction for multiple comparisons.}\\  
    \end{tabular}
\end{table*}

\section{Results}
\label{sec:results}
\subsection{Neurofibroma Segmentation Performance}

We evaluated six NF segmentation methods across three test sets to determine improvements over the baseline (Table \ref{tab:general segmentation performance}): a baseline 3D isotropic U-Net, a 3D anisotropic U-Net, a 3D anisotropic anatomy-informed U-Net, an ensemble of 3D anisotropic anatomy-informed U-Nets, and ensembles of 3D anisotropic anatomy-informed U-Nets with low- and high-confidence tumor candidate classification. The low-confidence threshold used for tumor candidate classification was set at the median of the lowest confidence values in the predicted NF segmentation masks, and the high-confidence threshold at a default value of 0.5. 

First, the segmentation metrics were calculated per scan. Methods incorporating anisotropy handling and anatomical knowledge outperformed the baseline across all test sets. Ensembling-based method further improved performance in datasets with higher variability in MRI protocols and tumor presentation (Test set \#2 and \#3). 
Tumor candidate classification had mixed results. The ensemble of the 3D anisotropic anatomy-informed U-Nets with low-confidence tumor candidate classification achieved the highest per-scan DSC of 0.64 on Test set \#1, representing a 68.4\% increase over the baseline. For Test set \#2 and \#3, the ensembling-based methods without and with high-confidence tumor candidate classification achieved the highest per-scan DSCs of 0.51 and 0.23, respectively.

Next, we evaluated per-tumor DSC for correctly detected tumors. Across all test sets, methods with anisotropic configurations, anatomical knowledge, and ensembling strategies consistently performed better than the baseline. However, the highest gains were observed with the ensemble of 3D anisotropic anatomy-informed U-Nets with high-confidence tumor candidate classification. This method achieved per-tumor DSC scores of 0.86, 0.86, and 0.89 on Test sets \#1, \#2, and \#3, showing significant improvements of 21.1\%, 14.7\%, and 4.7\% over the baseline, respectively. 

Based on these results, we selected the following methods for further analysis:
\begin{itemize}
    \item Baseline 3D isotropic U-Net (henceforth denoted as Method 1).
    \item 3D anisotropic U-Net (denoted as Method 2).
    \item Ensemble of 3D anisotropic anatomy-informed U-Nets (denoted as Method 3).
    \item Method 3 with high-confidence tumor candidate classification (denoted as Method 4).
\end{itemize}

\subsection{Neurofibroma Detection Performance}
\subsubsection{Per-Scan Analysis}
\label{subsubsec:per-scan detection performance}

We compared the tumor detection performance using per-scan F1 scores across three test sets (Table \ref{tab:per-scan detection performance}). Method 2 (i.e., anisotropic U-Net) significantly outperformed Method 1 (baseline) across all sets. Method 3 (with anatomical knowledge and ensembling) improved the results on Test set \#3 but did not consistently outperform Method 2 on the other sets. Method 4 (with tumor candidate classification) achieved the highest F1 scores of 0.44, 0.54, and 0.17 on Test set \#1, \#2, and \#3, corresponding to increases of 110\%, 170\%, and 1600\% over the baseline, although the improvement over Method 3 on Test set \#3 was not statistically significant.

\begin{table}
    \centering
\caption{Per-Scan F1 Scores for Tumor Detection Performance Across Three Test Sets. Methods: (1) 3D Isotropic U-Net, (2) 3D Anisotropic U-Net, (3) Ensemble of 3D Anisotropic Anatomy-Informed U-Nets, (4) Method 3 with High-Confidence Tumor Candidate Classification }
\label{tab:per-scan detection performance}
    \begin{tabular}{|>{\centering\arraybackslash}p{0.3\linewidth}|>{\centering\arraybackslash}m{0.1\linewidth}|>{\centering\arraybackslash}m{0.1\linewidth}|>{\centering\arraybackslash}m{0.1\linewidth}|>{\centering\arraybackslash}m{0.1\linewidth}|} \hline 
         Method $i$&  1&  2&  3& 4\\ \hline 
         \makecell[l]{\textbf{Test set \#1}} & \multicolumn{4}{c|}{}\\ \hline 
         \makecell{Per-scan F1 score \\ (Mean $\pm$ SD) $\uparrow$}&  0.21 $\pm$ 0.21&  0.40 $\pm$ 0.23&  0.40 $\pm$ 0.22& \textbf{0.44 $\pm$ 0.23}\\ \hline 
         \makecell{$p$-value \\ (Method $i$ vs $i-1$)}&  -&  $<0.01$&  0.52& 0.02
\\ \hline 
         \makecell[l]{\textbf{Test set \#2}} & \multicolumn{4}{c|}{}\\ \hline 
         \makecell{Per-scan F1 score \\ (Mean $\pm$ SD) $\uparrow$}&  0.20 $\pm$ 0.17&  0.46 $\pm$ 0.22&  0.48 $\pm$ 0.21& \textbf{0.54 $\pm$ 0.22}\\ \hline 
         \makecell{$p$-value \\ (Method $i$ vs $i-1$)}&  -&  $<0.01$&  0.97& $<0.01$\\ \hline 
         \makecell[l]{\textbf{Test set \#3}} & \multicolumn{4}{c|}{}\\ \hline 
 \makecell{Per-scan F1 score \\ (Mean $\pm$ SD) $\uparrow$}& 0.01 $\pm$ 0.01& 0.14 $\pm$ 0.12& \textbf{0.17 $\pm$ 0.14}&\textbf{0.17 $\pm$ 0.13}\\\hline
 \makecell{$p$-value \\ (Method $i$ vs $i-1$)}& -& $<0.01$& $<0.01$&0.52
\\\hline
\multicolumn{5}{p{230pt}}{SD – standard deviation. The highest metric values are in bold. Statistical significance was evaluated using the Wilcoxon signed-rank test for paired per-scan F1 score between consecutive methods (Method $i$ vs. Method $i-1$), with Bonferroni correction for multiple comparisons.}\\ 
    \end{tabular}
\end{table}

\subsubsection{Analysis Per Anatomical Region}
\label{subsec:per-region detection performance}
\begin{figure}
    \centering
    \includegraphics[width=1\linewidth]{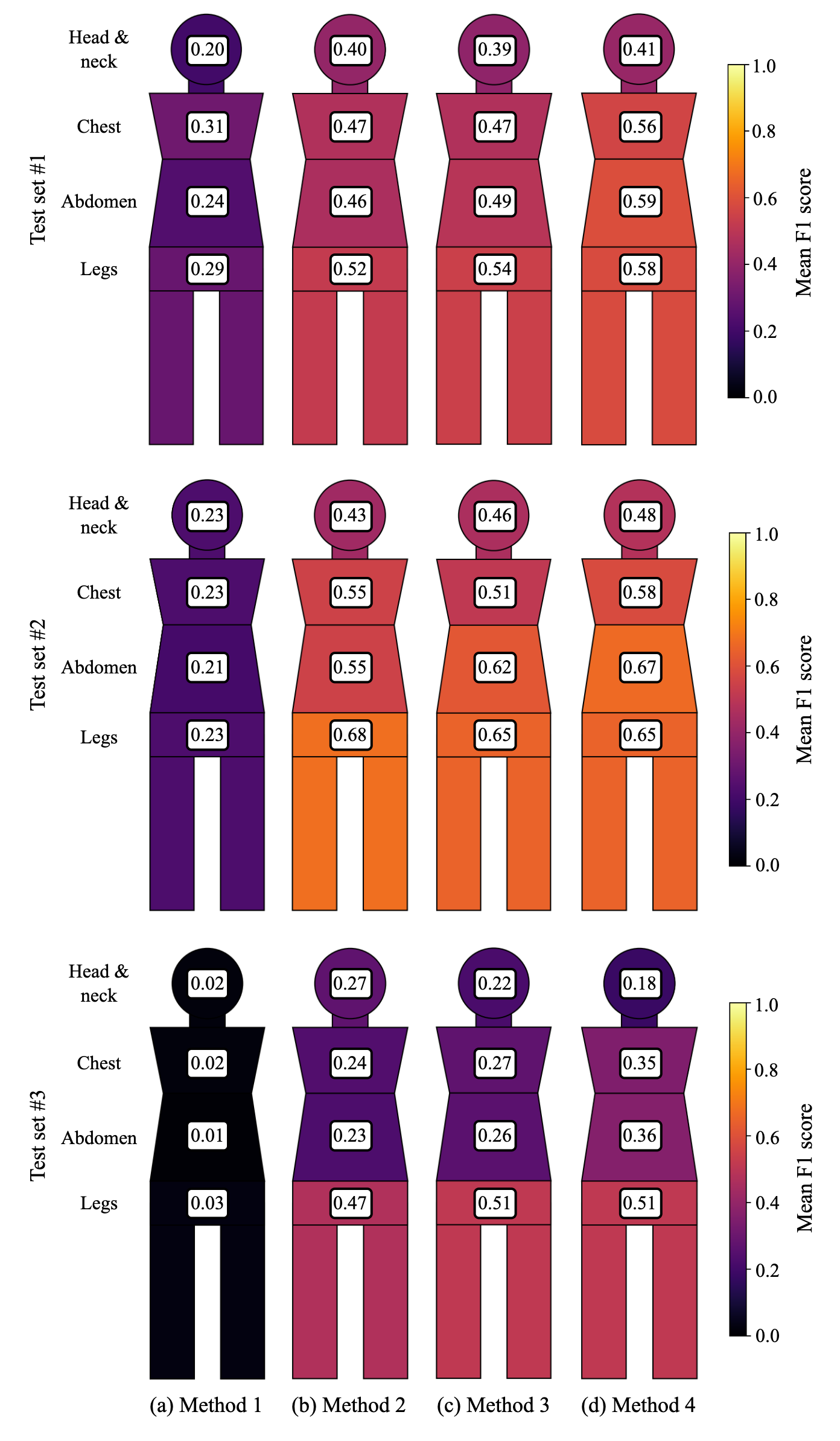}
    \caption{Mean F1 scores for tumor detection across anatomical regions for three test sets \#1, \#2, \#3 (rows), and four methods (columns): (a) Method 1 (3D isotropic U-Net), (b) Method 2 (3D anisotropic U-Net), (c) Method 3 (ensemble of 3D anisotropic anatomy-informed U-Nets), (d) Method 4 (Method 3 with tumor candidate classification).}
    \label{fig:per-region detection performance}
\end{figure}

We analyzed mean F1 scores for tumor detection across anatomical regions: head and neck, chest, abdomen, and legs (Fig. \ref{fig:per-region detection performance}). Performance generally improved with method complexity, from Method 1 to Method 4. Methods 2 (anisotropic U-Net) and 3 (with anatomical knowledge and ensembling) showed gradual improvements, whereas Method 4 (with tumor candidate classification) consistently outperformed the others in most regions. 

All methods performed best in the legs region, indicating fewer challenges for tumor detection in the lower body. The chest and abdomen regions showed moderate performance. The head and neck was the most challenging region for tumor detection, with the lowest mean F1 scores across all methods.

\subsection{Domain Shift Robustness Analysis}
\label{subsec:domain shift}
\begin{figure}
    \centering
    \includegraphics[width=1\linewidth]{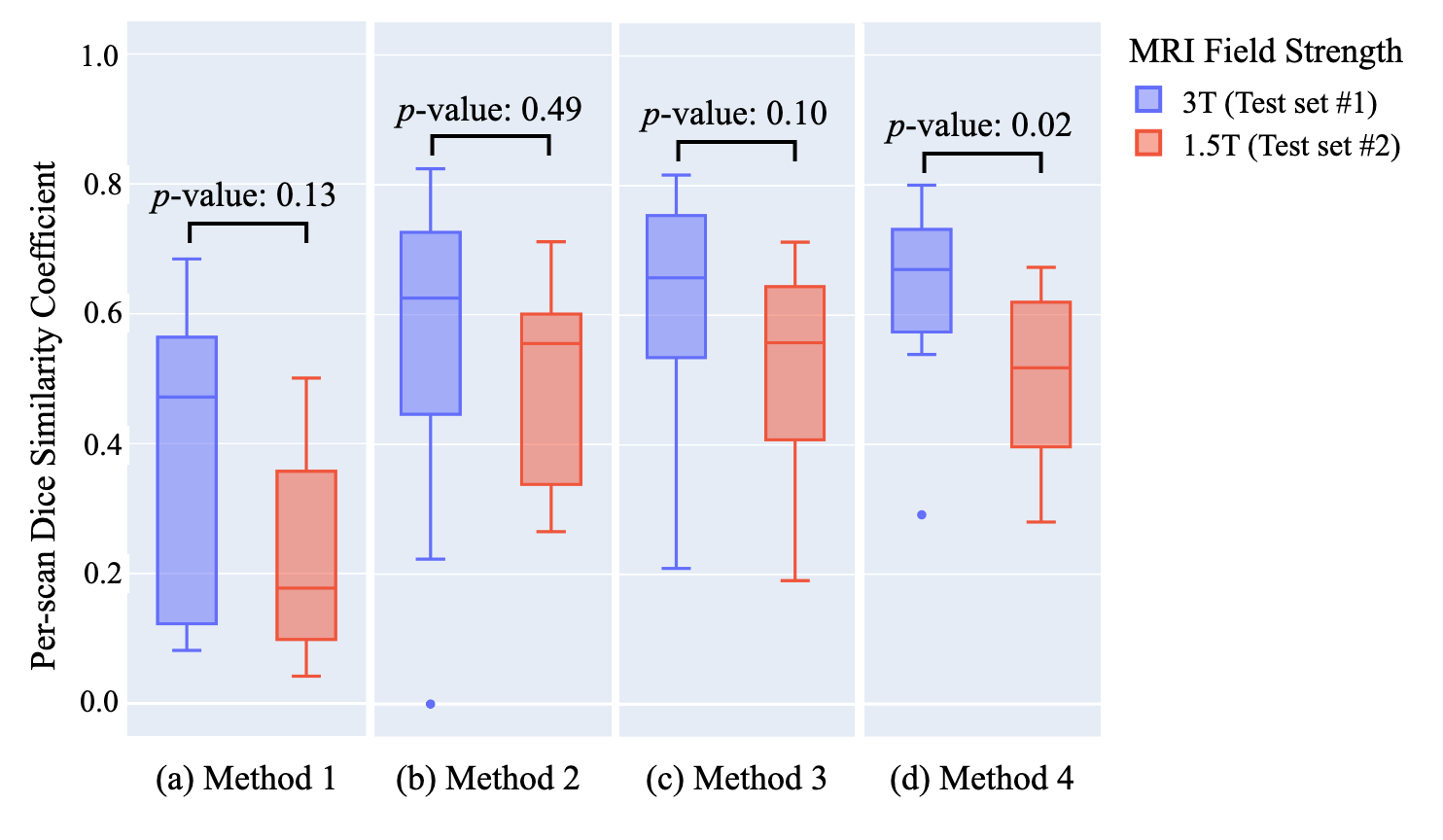}
    \caption{Degradation in neurofibroma segmentation performance (measured by per-scan Dice Similarity Coefficient) of methods tested on datasets with different MRI field strength (3T in Test set \#1 vs. 1.5T in Test set \#2), with corresponding p-values: (a) Method 1 (3D isotropic U-Net), (b) Method 2 (3D anisotropic U-Net), (c) Method 3 (ensemble of 3D anisotropic anatomy-informed U-Nets), (d) Method 4 (Method 3 with high-confidence tumor candidate classification).}
    \label{fig:domain shift}
\end{figure}

We assessed the robustness of NF segmentation methods to domain shifts by analyzing performance degradation when the neural networks trained on 3T MRI were tested on 1.5T MRI dataset (Test set \#2). Performance changes were measured using per-scan DSC (Fig. \ref{fig:domain shift}). The analysis revealed varying levels of robustness to domain shifts. Methods 1 (baseline), 2 (with anisotropy), and 3 (with anatomical knowledge and ensembling) experienced moderate performance differences, although these changes were not statistically significant. Method 4 (with tumor candidate classification) exhibited a statistically significant performance drop, indicating limited robustness to changes in MRI field strength compared to other methods.

\section{Discussion}
\label{sec:discussion}

\subsection{Performance Analysis of Neurofibroma Segmentation}
\label{subsec:performance analysis}

\subsubsection{Key Findings and Implications}
\begin{figure}
    \centering
    \includegraphics[width=1\linewidth]{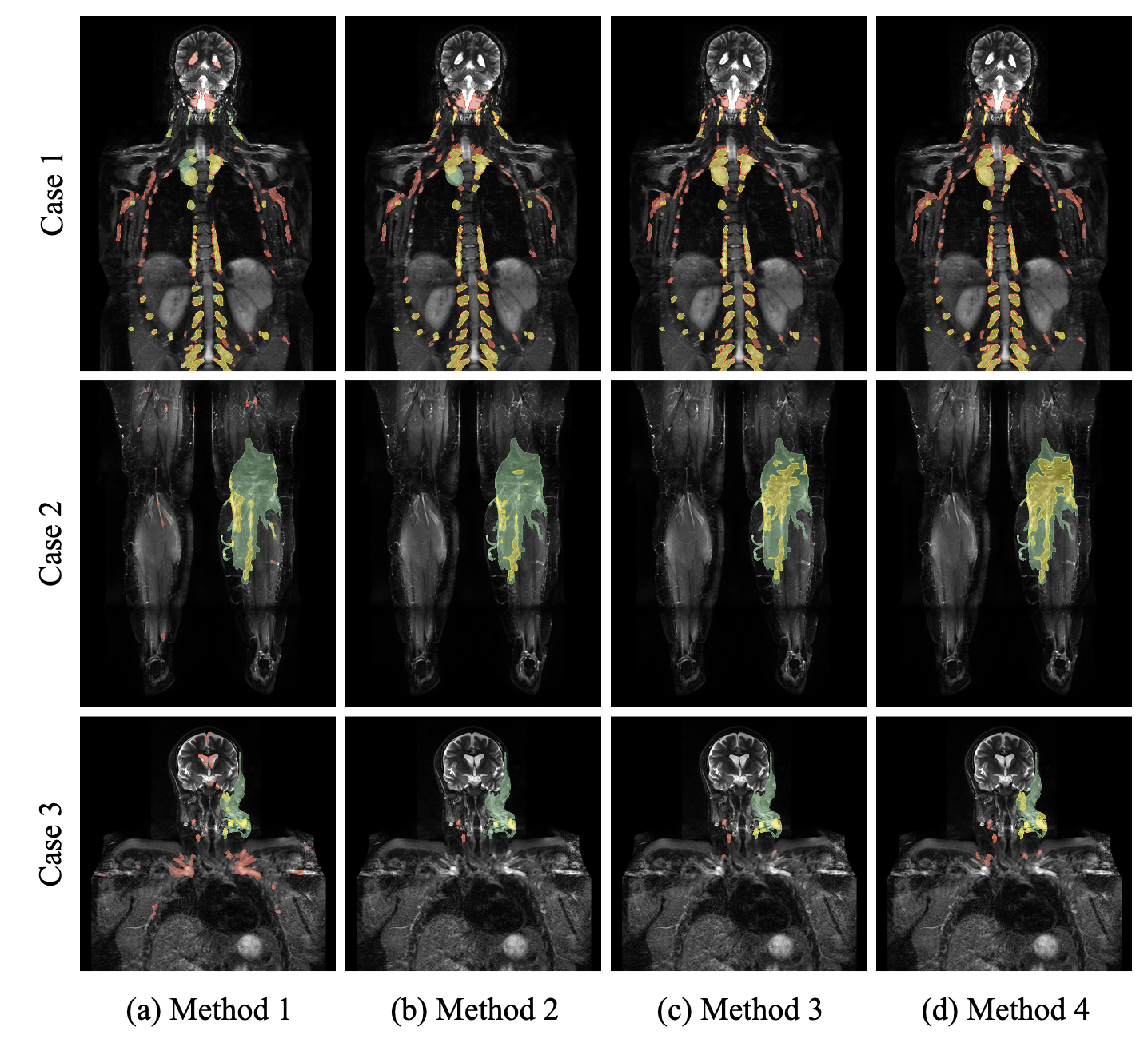}
    \caption{Comparison of neurofibroma segmentation results showing true positives (yellow), false positives (red), and false negatives (green) across three exemplar cases (rows) and four methods (columns): (a) Method 1 (3D isotropic U-Net), (b) Method 2 (3D anisotropic U-Net), (c) Method 3 (ensemble of 3D anisotropic anatomy-informed U-Nets), (d) Method 4 (Method 3 with high-confidence tumor candidate classification). Cases: (1) 25-year-old male with 626 tumors (3715.4 cm$^{3}$), (2) 55-year-old female with 4 tumors (734.7 cm$^{3}$), (3) 21-year-old female with 1 tumor (470.1 cm$^{3}$). }
    \label{fig:visual comparison}
\end{figure}

Introducing anisotropy into the U-Net architecture significantly improved NF segmentation, with the 3D anisotropic U-Net outperforming the baseline 3D isotropic U-Net, especially for hard-to-segment anatomical regions (e.g., improved segmentation in head, cases 1 and 3, Fig. \ref{fig:visual comparison}). The 3D anisotropic anatomy-informed U-Net enhanced performance by leveraging anatomical context to focus on likely NF regions, increasing true positive rates (e.g., improved segmentation in superior mediastinum, case 1, Fig. \ref{fig:visual comparison}). The ensemble strategy with tumor candidate classification achieved the best overall performance, reducing false positives by using radiomic features to distinguish tumors from non-tumor regions. It also improved segmentation in low-confidence areas, capturing larger tumor regions in diffuse NF cases (Fig. \ref{fig:visual comparison}, cases 2 and 3). However, this method was more sensitive to domain shifts.

\subsubsection{Comparison with Previous Studies}
Direct comparison with other studies is limited by data availability, but we benchmarked our findings across diverse datasets against \cite{zhang_dins_2022}, as both used the same evaluation protocol (Table \ref{tab:comparison with previous studies}). Compared to the nnU-Net baseline, our method showed superior performance on Test set \#1 and \#2, due to the integration of anatomical knowledge, ensembling, and tumor candidate classification, underscoring the potential of our approach for NF segmentation. Whereas the interactive DINs method performed well, our method achieved competitive performance on Test set \#1 with no manual intervention.

\begin{table}
    \centering
\caption{Comparison of the Ensemble of 3D Anisotropic Anatomy-Informed U-Nets with Tumor Candidate Classification (Method 4) against Previous Studies \cite{zhang_dins_2022}}
\label{tab:comparison with previous studies}
    \begin{tabular}{|c|c|c|c|c|} \hline 
         Method&  Dataset&  DSC $\uparrow$&  VOE $\downarrow$& ARVD $\downarrow$\\ \hline 
         Auto - DeepMedic&  DINs study&  0.06&  0.97& 77.95
\\ \hline 
         Auto - 3D U-Net&  DINs study&  0.06&  0.97& 77.62
\\ \hline 
         Auto - nnU-Net&  DINs study&  0.25&  0.84& 5.98
\\ \hline 
         Interactive - DINs&  DINs study&  0.69&  0.45& 0.64
\\ \hline 
         Method 4&  Test set \#1&  0.63&  0.53& 0.64
\\ \hline 
 Method 4& Test set \#2& 0.51& 0.64&1.12
\\ \hline 
 Method 4& Test set \#3& 0.23& 0.83&12.69
\\ \hline
\multicolumn{5}{p{230pt}}{DSC – Dice Similarity Coefficient, VOE – Volume Overlap Error, ARVD – Absolute Relative Volume Difference.}\\ 
    \end{tabular}
\end{table}

\subsubsection{Limitations and Further Directions}
Despite improvements, a key limitation of our method is sensitivity to domain shifts; a tumor candidate classifier trained on radiomic features from 3T MRI may not generalize well to 1.5T MRI data, emphasizing the need for feature harmonization. Performance also declines with sparse or small tumors (Test set \#3),  suggesting the integration of interactive segmentation for low tumor burden cases. Additionally, there is a need for standardized evaluation protocols in NF segmentation to enable consistent comparisons across studies.

\begin{table}
    \centering
\caption{Summary of Evaluation Protocols Used for Neurofibroma Segmentation Methods, Including Study Details and Metrics}
\label{tab:Evaluation_protocols}
    \begin{tabular}{|>{\centering\arraybackslash}p{0.2\linewidth}|>{\centering\arraybackslash}p{0.25\linewidth}|>{\raggedright\arraybackslash}p{0.4\linewidth}|} \hline 
         Study & Evaluation mode& Reported Metrics\\ \hline 
         Solomon \textit{et al.} \cite{solomon_automated_2004}, 2004  &Individual tumors within ROIs in 3D& • Correlation between volumes \newline • Coefficient of variation for repeatability assessment \newline • Segmentation time\\ \hline 
         Weizman \textit{et al.} \cite{weizman_interactive_2012}, 2012 &Individual / multiple tumors within ROIs in 3D& • Absolute volume error \newline • Overlap error \newline • Inter-rater variability \newline • Segmentation time \\ \hline 
         Weizman \textit{et al.} \cite{weizman_pnist_2014}, 2013 &Individual / multiple tumors within ROIs in 3D& • Absolute volume difference \newline • Volume overlap error \newline • 
 Absolute surface distance \newline • Intra-observer variability \newline • Correlation between methods \newline • Segmentation time\\ \hline 
         Pratt \textit{et al.} \cite{pratt_tumor_2015}, 2015 &Individual / multiple tumors within ROIs in 3D& • Volume overlap error \newline • Correlation between methods \newline • Intra- and inter-observer variability \newline • Segmentation time\\ \hline 
         Wu \textit{et al.} \cite{wu_deep_2020}, 2020 &Individual tumors within ROIs in 2D& • Intersection over Union \newline • Dice Similarity Coefficient \newline • Boundary F score \newline • Weighted Coverage \newline • F1 score \newline • Average Precision\\ \hline 
         Ho \textit{et al.} \cite{ho_image_2020}, 2020 &Individual tumors within ROIs in 3D& • Dice Similarity Coefficient \newline • Volume difference \newline • Segmentation time\\ \hline 
         Zhang \textit{et al.} \cite{zhang_dins_2022}, 2021 &All tumors in a scan in 3D& • Dice Similarity Coefficient \newline • Volumetric overlap error \newline • Absolute relative volume difference \newline • Number of required user interactions\\ \hline 
         Wu \textit{et al.} \cite{wu_dh-gac_2022}, 2022 &Individual / multiple tumors within ROIs in 2D& • Intersection over Union \newline • Dice Similarity Coefficient \newline • Boundary F \newline • Weighted Coverage\\ \hline 
\multicolumn{3}{p{200pt}}{ROIs – regions of interest.}
    \end{tabular}
\end{table}

\subsection{Limitations of Existing Evaluation Protocols}
\label{subsec:evaluation protocols discussion}

As outlined in Table \ref{tab:Evaluation_protocols}, multiple protocols and strategies exist for evaluating NF segmentation. Although recent studies adopted DSC to measure segmentation accuracy, DSC can be calculated differently depending on the study: in 2D or 3D, per tumor or over all tumors in a scan. Size thresholds for the inclusion of tumors in the evaluation also differ, with some studies using 75 cm$^{3}$ as a minimum \cite{solomon_automated_2004, weizman_interactive_2012}, whereas others include tumors as small as 5 cm$^{3}$ \cite{weizman_pnist_2014, pratt_tumor_2015, wu_deep_2020}. This lack of a unified evaluation protocol complicates direct comparisons between segmentation methods \cite{solomon_automated_2004, weizman_interactive_2012, pratt_tumor_2015, zhang_dins_2022}.

Per-scan DSC, commonly used in evaluation of DL-based methods, has limitations in the case of NF task, characterized by many small tumors spread across the body. Per-scan DSC is heavily skewed by background voxels in WB-MRI, and excluding the background increases sensitivity to small errors, resulting in high variability, especially in datasets with few tumors (e.g., Test set \#3).

In this regard, we evaluated the correlation between per-scan DSC and tumor burden and also examined the correlation between an alternative metric, per-tumor DSC, and respective tumor volume. Our analysis showed a moderate to strong positive correlation between per-scan DSC and tumor burden across all methods (Table \ref{tab:correlation}). In contrast, per-tumor DSC had a low correlation with tumor volume.

These findings suggest that while per-scan DSC reflects overall performance, it is not robust for evaluating segmentation of multiple small NFs. Thus, the evaluation protocol for NF segmentation remains an open research question.

\begin{table}
    \centering
\caption{Pearson Correlation Coefficients between Per-Scan and Per-Tumor Dice Similarity Coefficient with Per-Scan Tumor Burden and Tumor Volume Across Segmentation Methods: (1) 3D Isotropic U-Net, (2) 3D Anisotropic U-Net, (3) Ensemble of 3D Anisotropic Anatomy-Informed U-Nets, (4) Method 3 with High-Confidence Tumor Candidate Classification
}
\label{tab:correlation}
    \begin{tabular}{|>{\centering\arraybackslash}m{0.18\linewidth}|>{\centering\arraybackslash}m{0.13\linewidth}|>{\centering\arraybackslash}m{0.13\linewidth}|>{\centering\arraybackslash}m{0.13\linewidth}|>{\centering\arraybackslash}p{0.13\linewidth}|} \hline 
         Method $i$&  1&  2&  3& 4\\ \hline 
         Correlation: per-scan&  0.848&  0.614&  0.578& 0.603
\\ \hline 
         $p$-value: per-scan&  $<0.01$&  $<0.01$&  $<0.01$& $<0.01$\\ \hline 
         Correlation: per-tumor&  0.008&  0.034&  0.035& 0.033
\\ \hline 
         $p$-value: per-tumor&  0.550&  0.012&  0.011& 0.016
\\ \hline
    \end{tabular}    
\end{table}

\section{Conclusion}

We presented a novel automated segmentation pipeline for NF in T2w WB-MRI, incorporating anatomical knowledge through a 3D anisotropic U-Net with ensembling and radiomics-based tumor candidate classification. Our method outperformed reported fully automated methods like nnU-Net and achieved performance comparable to interactive, i.e. semi-automated methods such as DINs.

The integration into the 3D Slicer platform \cite{fedorov_3d_2012} showed promise for clinical application. However, challenges for clinical application remain, including handling variations in MRI field strength and low tumor burden cases. Our analysis also indicated that current evaluation metrics, like per-scan DSC, may not adequately capture the NF segmentation complexity, especially for multiple small tumors.

Future work aims to enhance robustness against domain shifts with MRI harmonization techniques, improve performance in low-burden cases, and explore integration with interactive segmentation methods for greater flexibility. We also advocate for the development of standardized evaluation frameworks to ensure consistent comparisons in NF segmentation research.

\section*{References}
\bibliographystyle{IEEEtran}
\bibliography{main}

\end{document}